\documentclass[aps,pra,twocolumn,showpacs,superscriptaddress]{revtex4}
\usepackage{times,amsmath,amssymb,amsbsy,epsfig,color,graphicx,multirow}

\begin{document}

\title{Hierarchical Gompertzian growth maps
with application in astrophysics}

\author{S. De Martino}
\affiliation{Dipartimento di Matematica e Informatica,
Universit\`a degli Studi di Salerno, Via Ponte don Melillo,
I-84084 Fisciano (SA), Italy} \affiliation{INFN Sezione di Napoli, Gruppo collegato di Salerno,
Fisciano (SA), Italy}

\author{S. De Siena}
\thanks{Corresponding Author. Electronic Address:
desiena@sa.infn.it}
\affiliation{Dipartimento di Matematica e Informatica,
Universit\`a degli Studi di Salerno, Via Ponte don Melillo,
I-84084 Fisciano (SA), Italy} \affiliation{INFN Sezione di Napoli, Gruppo collegato di
Salerno, Fisciano (SA), Italy}\affiliation{CNR-SPIN, Sede di Salerno, and CNISM,
Unit\`a  di Salerno, Fisciano (SA), Italy}


\begin{abstract}
The Gompertz model describes the growth in time
of the size of significant quantities
associated to a large number of systems, taking into account
nonlinearity features by a linear equation satisfied by
a nonlinear function of the size.
Following this scheme, we
introduce a class of hierarchical maps
which describe discrete sequences of intermediate characteristic scales.
We find the general solutions of the maps,
which account for a rich set of possible phenomena.
Eventually, we provide an important application, by showing
that a map belonging to the class so introduced
generates all the observed astrophysical length and mass scales.

\end{abstract}


\maketitle

The laws of growth of many systems, and the deep
origin of their characteristic scales of, e. g., length, mass, energy,
or numerosity,
are intensively investigated in many branches of
science, such as biomedicine, economy, and population dynamics.
The Gompertz model was originally introduced in 1825 by B. Gompertz
\cite{Gompertz} as a model
of human mortality: Gompertz found empirically
the fitting distribution of human age for a given community.
From the first half of the twentieth century
the Gompertz model, and the associated Gompertz equation,
have become a frequently used tool
to account for mechanisms of growth
\cite{Bajzer}.
The form of the Gompertz equation is:
\begin{equation}
z^{-1}\frac{d z}{dt} = \beta - \alpha \ln{(\frac{z}{{\tilde z}}}),
\label{Gompertzeq1}
\end{equation}
where $z$
describes the "size" (not necessarily a spatial size) of
some quantity
characterizing the system,
$\beta$ and $\alpha$ denote two positive constants
with the dimensions of the
inverse of time, and ${\tilde z}$ is a
constant with the same dimensions of $z$.
Eq. (\ref{Gompertzeq1}) can be recast as:
\begin{equation}
\frac{d (\ln{s})}{dt} = - \alpha \ln{s},
\label{Gompertzeq2}
\end{equation}
where $s(t) \doteq z(t)/z_{\infty}$, and
$z_{\infty} = {\tilde z} \exp{(\beta/\alpha)}$.
The Gompertz equation is then characterized by four
parameters (all dependent on the specific system):
$\alpha$, $\beta$, $z_{\infty}$,
and the initial condition ("scale") $z(0) = z_0$.
Its solution is:
\begin{equation}
z(t) = z_{\infty} \exp{[(\ln\gamma) \cdot e^{-\alpha t}]},
\label{Gompertzsol}
\end{equation}
where $\gamma \doteq z_0/z_\infty$.
It is immediately verified that this solution
always approaches monotonically in time $z_\infty$.
Depending on the conditions
$z_0 < z_\infty \equiv {\tilde z} \exp{\beta/\alpha} \; \; \;  (\gamma <1)$,
or $z_0 > z_\infty \equiv {\tilde z} \exp{\beta/\alpha} \; \; (\gamma > 1)$,
the system  monotonically grows or monotonically
decreases, respectively, from the dimension $z_0$ to
the dimension $z_\infty$, approaching the asymptotic value
$z_\infty$ with the characteristic time $\alpha^{-1}$.
In TABLE I we list the main symbols which have been exploited,
or which will be exploited in the following, with their meanings.

\begin{table}[h!]
  \begin{center}
\begin{tabular}{|c|c|}
  \hline
  Symbol & Meaning \\ \hline
  $\alpha, \beta$  & Parameters of the Gompertz eq. \\ \hline
  $z(t)$ & Size at time $t$ \\ \hline
  $z_0$ & Initial size \\ \hline
  $z_{\infty}$ & Asymptotic (or reference) size \\ \hline
  $\gamma$ & Ratio $z_0/z_{\infty}$ \\ \hline
  $s(t)$ & Relative size $z(t)/z_{\infty}$ \\ \hline
  $s_n$ & $n$-th relative size $z_n/z_{\infty}$ in the discrete map \\ \hline
  $y_n$ & $\ln{s_n}$ \\ \hline
  ${\tilde \alpha}$ & Parameter of the discrete map \\ \hline
  $R_n$ & $n$-th astrophysical length scale \\ \hline
  $R$ & Max. astroph. length scale (Radius of the obs. univ.)\\ \hline
  $\lambda$ & Min. length scale $R_0$ (radius of a nucleon) \\ \hline
  $M_n$ & $n$-th astrophysical Mass scale \\ \hline
  $M$ & Max. astroph. mass scale (Mass of the obs. univ.)  \\ \hline
  $m$ & Min. mass scale $M_0$ (mass of a nucleon)\\ \hline
  $c$ & Velocity of the light \\ \hline
  $G$ & Gravitational constant \\
  \hline
\end{tabular}
\end{center}
  \caption{Meaning of the symbols}
\end{table}

We observe that the Gompertz model includes
in a very interesting and
peculiar way nonlinearity, which is a macroscopic
mirror of a nonlinear microscopic background ruling
the growth of natural systems. In fact, the equation
looks linear in a suitable function $y(s)$
of the original (relative) size $s$, while the nonlinearity
is introduced by the fact that the function $y$
is itself a nonlinear function of $s$;
in particular, $y(s) = \ln{s}$,
or, equivalently, $s(y) = \exp{y}$.
On the other hand, while the Gompertz equation
describes a continuous growth of a size in time,
we know that many systems
are placed on some
discrete sequence of sizes, ranging from
a minimum to a maximum scale.
Therefore, following the Gompertz model,
we try to describe these sequence
of scales by a sequence $\{s_n\}$
of relative sizes resulting from
the solution of a {\it linear} map
fulfilled by a {\it nonlinear}
function, $y(s_n) \equiv y_n$, of $s_n$:
\begin{equation}
y_{n+1} = \delta_n y_n,
\label{Gendiscretemapy}
\end{equation}
where $\delta_n$ are proportionality coefficients.
The solution of the map is obviously:
\begin{equation}
y_n = y_0 \prod_{k=0}^{n} \delta_k.
\label{GendiscretemapySol}
\end{equation}
Here, $s_n$ will be defined by $s_n \doteq z_n/z_{\infty}$,
where $z_n$ is the $n$-th size, and $z_{\infty}$
is an asymptotic, or reference, size. Moreover, for
further generalization, we at first allow the
dependence of the proportionality constant
on $n$ (analogously, we could
promote $\alpha$ to a suitable
function of the time in the Gompertz equation).
The choice of the
nonlinear function $y(s_n) \equiv y_n$
is obviously a crucial question: this choice,
in principle,
will depend on the specific system, and it
should be made following
physical criteria and
phenomenological observations.
But the Gompertz model suggests that
the logarithm  can play
a privileged and somewhat universal role.
This suggestion is strengthened by
the connection among Gompertz model,
lognormality and stability, as we will briefly discuss in
the conclusions \cite{NoiStochGomp}.
We then select $y_n = \ln{s_n}$ (or
$s_n = \exp{y_n}$). Thus, if $\gamma$
is defined as in the Gompertz equation,
$y_0 = \ln{\gamma}$, and
the solution (\ref{GendiscretemapySol})
gives:
\begin{equation}
s_{n+1} = {s_n}^{\delta_n},
\label{Gendiscretemaps}
\end{equation}
i. e.
\begin{equation}
s_{n} = ({\gamma})^{\prod_{k=0}^{n} \delta_k}.
\label{GendiscretemapsSol}
\end{equation}
Furthermore, resorting again to the Gompertz
model, we make the simplifying choice
$\delta_n \equiv \delta \equiv 1 - {\tilde \alpha}$,
obtaining
\begin{equation}
y_{n+1} = (1 - {\tilde \alpha}) y_n ,
\label{discreteGmapy},
\end{equation}
\begin{equation}
y_{n} = (1 - {\tilde \alpha})^{n} \ln{\gamma} ,
\label{discreteGmapysol},
\end{equation}
\begin{equation}
s_{n+1} = s_n^{(1 - {\tilde \alpha})},
\label{discreteGmaps}
\end{equation}
and
\begin{equation}
s_n = \gamma^{(1 - {\tilde \alpha})^n}.
\label{discreteGmapssol}
\end{equation}
In fact, Eq. (\ref{discreteGmapy}) looks as the form of the discretized
Gompertz equation. It is, however, worth to be remarked
that the physical meaning of the discrete map
is different from that of the continuous equation; in
particular, we are not necessarily authorized to interpret
Eq. (\ref{discreteGmapy}) as an equation in (discretized) time.

We now comment on the behavior of the solutions
(\ref{discreteGmapssol}).
We see that a monotone sequence (increasing if ${\gamma} < 1$,
decreasing if $\gamma > 1$) is assured only if ${\tilde \alpha} < 1$
(while monotonicity is always assured by the
Gompertz equation).
If, instead, ${\tilde \alpha} > 1$, the map (\ref{discreteGmapy}),
(with solution(\ref{discreteGmapysol}))
changes its sign at each step, and, correspondingly,
the values of $z_n$ {\it oscillate}, but with two further
possible behaviors. If ${\tilde \alpha}$ is not an integer number,
the size of the system approaches the asymptotic value $z_{\infty}$ by
oscillating in alternating way above and below it,
with more and more damped oscillations.
If ${\tilde \alpha}$ is an integer number,
the size of the system oscillates indefinitely
above and below the reference value $z_{\infty}$,
assuming in alternating way the pair of values
$\gamma^{({\tilde \alpha}-1)}$
and $\gamma^{(1-{\tilde \alpha})}$,
and does not converge to any asymptotic
limit.
Therefore, we see that the discrete maps
so introduced describe a variety of
situations.

In order to investigate the usefulness
of the maps (\ref{discreteGmaps}), (\ref{discreteGmapssol}),
we move towards an application, by premising some
considerations. We observe that,
if the growth is monotone (${\tilde \alpha} < 1$),
usually one knows (phenomenologically)
the initial and the final sizes ($z_0, z_{\infty}$)
of a system, and that,
if $\gamma < 1$
(aggregation process) the initial value $z_0$ is noting but the
size
of the more elementary constituent of the system,
and $z_{\infty}$ is the
maximum size, while if
$\gamma > 1$
(fragmentation process), $z_{\infty}$ is the size
of the more elementary constituent, and $z_0$ is
instead the maximum size. The (phenomenological) knowledge
of the extremal sizes, together with a good fit for
the value of ${\tilde \alpha}$, is sufficient
to determine all the intermediate scales.
Obviously, a more ambitious goal could be to
obtain the hierarchical sequence of characteristic scales
$z_n$ by fixing phenomenologically only
the initial size $z_0$, and obtaining the values of ${\tilde \alpha}$
and of $z_{\infty}$ by independent theoretical hypotheses.
This, as we will show, can be done when the physical background
underlying the system is well established.

Now, we show that Eq. (\ref{discreteGmapssol})
can be applied in the framework of astrophysics:
we assume that the components
on different scales of the observed universe
organize themselves according to
our scheme, and proceed to verify this hypothesis. We choose
for $z_n$ the sequence of the length scales ("radii")
of the astrophysical aggregations, and replace the symbols
$z_n$ with $R_n$ and $z_{\infty}$ with $R$,
and all the others in a consistent manner. Moreover, we fix ${\tilde \alpha} = 1/2$.
Being ${\tilde \alpha} < 1$, the map is monotone. We assume also
that $R_0$ is the size of the elementary constituent (i. e. the minimum size).
Thus, the radii $R_n$ are monotonically increasing, and their
asymptotic limit $R$ is the maximum radius (i. e. the observed radius of
the universe). We remark that in our computation we consider only the
order of magnitude of the radii, expressed (in centimeters) as powers of $10$.
It is natural to choose as elementary constituent a nucleon, because
the system is ruled by gravitation, which in turn is
determined by the mass distribution; and
nucleons contain all the significant (observed) mass.
Our minimum size is then the radius
of a nucleon as measured by $\alpha$-particles scattering:
$R_0 \equiv \lambda = 10^{-13} cm$.
It is also known that $R$, the observed radius of the universe, is
$R = 10^{26} cm$ \cite{Weinberg}. Then $\gamma = 10^{- 39}$. The map (\ref{discreteGmaps}),
with ${\tilde \alpha} = 1/2$, becomes a {\it square-root map},
while if we use $s_n = R_n/R$ we obtain for the radii a {\it geometric-mean map}:
\begin{equation}
R_{n+1} = (R_n \; R)^{\frac{1}{2}}.
\label{GeomMeanMap}
\end{equation}
Now, exploiting Eq. (\ref{discreteGmapssol}) with $\gamma = 10^{- 39}$ and
${\tilde \alpha} = 1/2$, or, equivalently, Eq. (\ref{GeomMeanMap}),
we obtain the rapidly convergent sequence in the first column of TABLE II.

\begin{table}[h!]
  \begin{center}
\begin{tabular}{|c|c|c|}
  \hline
     RADIUS    &    MASS      &  AGGREGATE \\ \hline
  $10^{6} cm$  & $10^{15} g$  & planetesimal \\ \hline
  $10^{6} cm$  & $10^{34} g$  & neutron star \\ \hline
  $10^{16} cm$ & $10^{34} g$  & solar system \\ \hline
  $10^{16} cm$ & $10^{44} g$  & s.m. black hole \\ \hline
  $10^{21} cm$ & $10^{44} g$  & typical galaxy \\ \hline
  $10^{23} cm$ & $10^{49} g$  & cluster of galaxies \\ \hline
  $10^{25} cm$ &  $10^{52} g$ & supercluster of galaxies \\ \hline
  $10^{26} cm$ & $10^{54} g$  & observed universe \\
  \hline
\end{tabular}
\end{center}
  \caption{List of astrophysical length scales and of the corresponding mass scales,
  with the associated astrophysical aggregates}
\end{table}

These length scales represent, in order of magnitude,
just the main six observed astrophysical length scales \cite{Weinberg}
(the presence of two aggregates both for radius $R_1$ and
for radius $R_2$ is discussed later).
Therefore, we have obtained a first
important result: we can fit
the intermediate astrophysical length scales by
choosing a specific value
of ${\tilde \alpha}$, and by exploiting the
experimentally determined values of the two
extremal length scales.

However, as remarked previously,
a scientific theory with some {\it predictive}
power should be able to determine,
on the basis of few other
assumptions, both the (apparently
arbitrary) choice ${\tilde \alpha} =1/2$, and
the value of the maximum size $R$, by the
mere phenomenological knowledge of the
minimum scale.
We now show that this is actually possible
by resorting to a suitable assumption naturally
suggested by
the physics of the gravitational systems:
{\it We assume that the dimension of
the aggregate with radius $R_1$ (and mass ${\bf M}$)
is given by the Schwartzschild radius of a
three-dimensional close packing of elementary constituents
(nucleons: radius $\lambda$, mass $m$), i. e.
by a "collaps"
restricted by the constraint that
the escape velocity is the maximum one: the velocity
$c$ of the light}. This hypothesis leads to
the following two conditions:
\begin{equation}
\frac{{\bf M}}{R^{3}_1} = \frac{m}{\lambda^{3}}; \; \; \; \; G \frac{{\bf M}}{R_1} = c^2,
\label{FirstAggregate}
\end{equation}
where $G$ is the gravitational constant.
The second equation is the Schwartzschild condition for the radius,
while the first one
requires that the mass density per unit volume of the first aggregate
coincide with that of a nucleon (three-dimensional
close packing). By eliminating the mass $M$, we obtain:
\begin{equation}
R_1 = (\lambda R)^{\frac{1}{2}},
\label{FirstAggregateRadius}
\end{equation}
with $R = (\lambda c)^{2}/Gm$.
Inserting the numerical values of $\lambda, c, g, m$
provides $R = 10^{26} cm$, i. e. just the
observed radius of the universe. Thus, we see that our simple hypothesis
leads both to an independent determination of the
maximum size $R$, {\it and} to the value of ${\tilde \alpha}$
which coincides with our original choice
(Note in fact that Eq. (\ref{FirstAggregateRadius}) is Eq. (\ref{GeomMeanMap})
with $n = 0$). Here we comment also
that, by defining (in order of magnitude) the age $T$
of the observed universe
by $R = c T$, from the expression soon established for $R$
we get: $T = \lambda^2 c/Gm \equiv 10^{16} s$,
i. e. just the right order of magnitude.

Finally, we aim to determine, besides
the length scales, also the masses of the astrophysical aggregates.
To this purpose, we note that, if some other quantity of the system,
say $M_n$, is connected to $z_n$ by the "allometric" relation:
$M_n = b z_{n}^{\delta}$, then the quantity $q_n \doteq M_n/M_{\infty}$ satisfies
the same map as $s_n$, and has the same form of
solution with the replacement
$\gamma \rightarrow \gamma^{'} = \gamma^{\delta}$:
\begin{equation}
q_n = (\gamma^{'})^{(1 - {\tilde \alpha})^n} \equiv \gamma^{\delta (1 - {\tilde \alpha})^n}.
\label{discreteGmapMsol}
\end{equation}
Moreover, putting together the allometric relations at a generic $n$ and at
$n=0$, we obtain also the proportionality constant $b$ as:
$b = M_0/z^{\delta}_0$.

In the case of the universe, we consider the sequence
of the masses $M_n$, and the sequence of their radii $R_n$,
and find their "allometric" relation
by introducing another assumption (of "minimum fluctuation"),
probably less intuitive
with respect to the first one, but reasonable
if one considers that the very complex
character of the gravitational systems,
and the huge numbers of elementary constituents
induce instabilities and fluctuations:
{\it An aggregate can bind a test particle up to a distance
where the gravitational force generated by the aggregate
on a test particle
reduces to a value comparable with the background fluctuating
force, whose magnitude is determined by the elementary
constituent}. The claim gives, for each $n$:
\begin{equation}
G \frac{M_n}{R^{2}_n} = G \frac{m}{\lambda^2},
\label{fluctuations}
\end{equation}
i. e.:
\begin{equation}
M_n = b {R^{2}_n},
\label{GravAllometry}
\end{equation}
where $b = m/\lambda^2 = 100 g/cm^2$.
Then, $\delta = 2$,
and, being ${\tilde \alpha =1/2}$, we obtain
the sequence:
\begin{equation}
M_n = \gamma^{2(2^{-n} - 1)} m,
\label{GravMassSol}
\end{equation}
where, by denoting
$M$ (instead of
$M_{\infty}$) the asymptotic value providing
the total mass of the observed universe,
we have exploited: $m/M = \gamma^2$.
Then, once the sequence for the radii is established,
the allometric relation, and the knowlege
of the minimum mass $m$, automatically
provide the sequence of the intermediate mass
scales until the maximum one.
From $\gamma = 10^{- 39}$ and $m = 10^{- 24} \; g$,
we obtain (in order of magnitude)
the sequence of the second column in TABLE II, which
represents just the sequence of the
observed astrophysical masses.
In the last column of TABLE II
we describe the corresponding structures.
Note also that the total mass of the observed universe
corresponds to that of $\gamma^{-2} \equiv 10^{78}$ nucleons, in perfect
agreement with the central value obtained from nucleosynthesis
calculations \cite{Weinberg}.

\vspace{0.3cm}

\noindent We conclude this letter with the following
observations and remarks:

Being the macroscopic
behavior of complex systems
the result of the collective
behavior of a huge number of elementary constituents, it
must be a mean effect deriving by
a suitable microscopic (stochastic) model.
In fact, in
ref. \cite{NoiStochGomp} it has been recently proved that
the Gompertz equation is a {\it macroscopic consequence} of the action of
a microscopic, lognormally distributed diffusion process performed
by the elementary constituents of the system,
because this equation holds for the
{\it median} of the process,
which then provides
the evolution in time of
the macroscopic characteristic size.

We remark that hypotheses and results on the astrophysical structures
here discussed, and the hierarchical maps for the sequence of
astrophysical radii and masses
are contained in papers previously published
by the authors in
collaboration with other researchers \cite{NoiUniverso}.
However, we as well remark that in this letter we
frame these results in a model
based on discrete maps
deduced by the Gompertz model
of growth. The background stochastic origin of the
Gompertz model \cite{NoiStochGomp}, together with the
derivation from physical hypothesis
of the values of ${\tilde \alpha}$ and of
the asymptotic length $R$,
lead us to interpret our scheme as
a first important step towards
the discovery of the underlying process performed
by the elementary constituents, and
responsible for
the  observed sequence of astrophysical scales.
Note, in fact, that it clearly
emerges from our model the role
of a geometrical progression, whose relevance
in the framework of natural phenomena and
in a variety of experimental environments
was highlighted already at the end of nineteenth century
by F. Galton \cite{Galton} and
by D. McAlister
\cite{McAlister}, which
showed that the geometrical mean (median)
describes the behavior of a large set
of natural phenomena
better than the arithmetic one.

The discrete maps here introduced
can provide a general model that can be potentially
exploited to search for the enlightenment of a large number
of growth phenomena in many fields of research.
Namely searching for sequences of significant
scales associated to different systems, which possibly
develop these intermediate scales step by step in time,
or which organize themselves on all these scales simultaneously.
It is also of great interest to
investigate the existence of natural quantities whose evolution
is characterized by oscillations,
as described by the maps with ${\tilde \alpha} >1$.

We remark that the apparent
discrepancy between the first relation in Eq.
(\ref{FirstAggregate}) and the relation
(\ref{fluctuations})
is solved by the fact that to the same radius $R_n$
can be associated the mass $M_n$
{\it and also} the mass $M_{n+1}$, where the mass $M_n$
satisfies the relation (\ref{fluctuations})
(two-dimensional close packing), while the mass
$M_{n+1}$  satisfies the relation (\ref{FirstAggregate})
(three-dimensional close packing, "critical" aggregate)
\cite{NoiUniverso}.
For example, as is well known, a neutron star (three dimensional
close packing with mass $M_2 = 10^{34} g$) has, in order of magnitude,
the same radius ($R_1 = 10^{6} m$)
of a planetesimal (two dimensional
close packing with mass $M_1 = 10^{15} g$).
Similar considerations can be
extended by comparing a typical galaxy
with a supermassive black-hole at the center
of the galaxy bulge, whose estimated mass
(until billions of solar masses, i. e.
$10^{42} g$)
is in fact comparable with that of the whole galaxy
\cite{smbh}.
This explains the two structures
associated to the first two radii in Table I.
Summing up, our model
allows two {\it possible} sequences, given by:
$\{R_n, M_n\}$ (two-dimensional close-packed aggregates)
and $\{R_n, M_{n+1}\}$
(three-dimensional close-packed, critical aggregates),
and this scheme is confirmed at least
on scale $R_1$, and on scale $R_2$.

A question which is worth to be
deepened is the meaning of the proportionality parameters
in the linearized maps. But this question is connected
to the more general, and already raised, question
of the possible stochastic background underlying the maps,
and will be addressed in forthcoming papers.

\end{document}